\documentclass[reqno]{amsart}
\usepackage{amssymb}

\newtheorem{theorem}{Theorem}[section]

\newtheorem{corollary}[theorem]{Corollary}

\theoremstyle{definition}

\theoremstyle{remark}

\numberwithin{equation}{section}

\begin{document}

\title[Equilibrium configuration of the $BC$-type RS system]
{On the Equilibrium Configuration of the BC-type Ruijsenaars-Schneider System}

\author{J.F. van Diejen}
\address{Instituto de Matem\'atica y F\'{\i}sica, Universidad de Talca,
Casilla 747, Talca, Chile}

\date{August, 2004}

\begin{abstract}
\noindent It is shown that the ground-state equilibrium configurations of the trigonometric
$BC$-type Ruijsenaars-Schneider systems are given by the zeros of
Askey-Wilson polynomials.
\end{abstract}

\maketitle

\section{Introduction}\label{sec1}
The Ruijsenaars-Schneider systems \cite{rui-sch:new,die:integrability,die:deformations}
are integrable deformations of
the celebrated Calogero-Moser $n$-particle models \cite{ols-per:classical}.
It is well-known that the equilibrium configurations of the
Calogero-Moser models are described by
the zeros of classical orthogonal polynomials such as the Hermite, Laguerre, Chebyshev, and
Jacobi polynomials \cite{cal:zeros,cal:equilibrium,cal-per:properties,ols-per:classical}.
This connection between the equilibria of one-dimensional integrable particle models and the locations of zeros of the classical hypergeometric orthogonal polynomials, first observed by Calogero, is closely related to a beautiful electrostatic interpretation of the zeros of orthogonal polynomials due to Stieltjes \cite{sze:orthogonal}.
Recently, it was noticed that the equilibrium configurations of the Ruijsenaars-Schneider systems
can also be described in a similar way by means of the zeros orthogonal polynomials
\cite{rui:action,rag-sas:quantum,oda-sas:equilibria}; all polynomials that appear in
in this context turn out to be classical in the sense that they sit somewhere in Askey's hierarchy of (basic) hypergeometric orthogonal polynomials \cite{ask-wil:some,koe-swa:askey}.
The top of this hierarchy is formed by the
Askey-Wilson polynomials \cite{ask-wil:some}. (All other (basic) hypergeometric families of classical orthogonal polynomials are special (limiting) cases of the Askey-Wilson polynomials
\cite{koe-swa:askey}.) In this note we show that the zeros of these Askey-Wilson
polynomials correspond to the ground-state equilibrium configurations of the trigonometric $BC$-type Ruijsenaars-Schneider systems introduced in Refs. \cite{die:integrability,die:deformations}. In the rational limit, one recovers the characterization of the ground-state equilibrium configurations of the rational $BC$-type Ruijsenaars-Schneider systems in terms of Wilson polynomials \cite{wil:some} due to Odake and Sasaki \cite{oda-sas:equilibria}

We will need to employ the following standard conventions from the theory of
(basic) hypergeometric orthogonal polynomials \cite{ask-wil:some,koe-swa:askey}:
$q$-shifted factorials are denoted by
\begin{equation*}
(a;q)_k:=\begin{cases}
1 &\text{for}\; k=0, \\
(1-a)(1-aq)\cdots (1-aq^{k-1}) &\text{for}\; k=1,2,3,\ldots ,
\end{cases}
\end{equation*}
with the convention that $(a;q)_\infty:=\prod_{k=0}^\infty
(1-aq^k)$ (for $|q|<1$); products of $q$-shifted factorials are
abbreviated in the usual way via
\begin{equation*}
(a_1,\ldots ,a_{r};q)_k:=(a_1;q)_k\cdots (a_r;q)_k.
\end{equation*}
The ${}_{r+1}\Phi_r$ terminating
basic hypergeometric series is defined as
\begin{equation*}
{}_{r+1}\Phi_r \left[ \begin{array}{c} q^{-n},a_1\ldots ,a_{r} \\
b_1,\ldots ,b_r  \end{array}  \mid q;z  \right] :=
\sum_{k=0}^n
\frac{(q^{-n},a_1,\ldots ,a_{r};q)_k}{(q,b_1,\ldots ,b_{r};q)_k} z^k
\end{equation*}
(where it is assumed that the parameters are such that denominators
do not vanish).
The hypergeometric degeneration of this series is given by
\begin{equation*}
{}_{r+1}F_r \left[ \begin{array}{c} -n,a_1\ldots ,a_{r} \\
b_1,\ldots ,b_r  \end{array}  \mid z  \right] :=
\sum_{k=0}^n
\frac{(-n,a_1,\ldots ,a_{r})_k}{(1,b_1,\ldots ,b_{r};q)_k} z^k,
\end{equation*}
where $(a_1,\ldots ,a_{r})_k:=(a_1)_k\cdots (a_r)_k$
with $(a)_k:=a(a+1)\cdots (a+k-1)$ (and $(a)_0:=1$ by convention).

\section{The trigonometric $BC$-type Ruijsenaars-Schneider system}\label{sec2}
The trigonometric $BC$-type Ruijsenaars-Schneider system is a one-dimensional
$n$-particle model characterized by the
Hamiltonian \cite{die:integrability,die:deformations}
\begin{subequations}
\begin{equation}\label{h1}
H(\mathbf{p},\mathbf{x})=
\sum_{j=1}^n \left(
\cosh (p_j)\sqrt{V_j(\mathbf{x})V_j(-\mathbf{x})}
- \bigl( V_j(\mathbf{x})+V_j(-\mathbf{x})\bigr) /2\right) ,
\end{equation}
where
\begin{gather}
\label{h2} V_j(\mathbf{x}) = w(x_j) \prod_{1\leq k\leq n,\, k\neq j} v(x_j+x_k)\, v(x_j-x_k)  ,\\
\label{h3} v(x)= \frac{\sin (x+ig)}{\sin (x)},\quad
w(x)= \frac{ \sin (x+ig_1)\cos (x+ig_2 )\sin (x+ig_3)\cos(x+ig_4)}{\sin^2 (x)\cos^2(x)}
\end{gather}
\end{subequations}
(and $i:=\sqrt{-1}$).
Throughout this note we will assume that the coupling parameters
$g$ and $  g_r $ ($r=1,2,3,4$) are {\it positive}. This guarantees in particular
that  the Hamiltonian $H(\mathbf{p},\mathbf{x})$
\eqref{h1}-\eqref{h3} constitutes a nonnegative (smooth) function
on the phase space
\begin{equation}\label{phase-space}
\Omega = \{ (\mathbf{p},\mathbf{x})\in \mathbb{R}^{2n}\mid
0< x_1 <x_2<\cdots < x_{n-1}<x_n < \pi /2 \} .
\end{equation}
Indeed, one has that
$H(\mathbf{p},\mathbf{x})\geq H(\mathbf{0},\mathbf{x})\geq 0$
(since $V_j(-\mathbf{x})=\overline{V_j(\mathbf{x})}$ and
$|V_j(\mathbf{x})| \geq \text{Re} (V_j(\mathbf{x}))$).

\section{The ground-state equilibrium configuration}\label{sec3}
The equilibrium configurations correspond to the critical points of the
Hamiltonian $H(\mathbf{p},\mathbf{x})$ and the ground-state equilibrium
configurations correspond in turn to the global minima.
It is clear that the only way in which the nonnegative
$BC$-type Ruijsenaars-Schneider
Hamiltonian $H(\mathbf{p},\mathbf{x})$
\eqref{h1}-\eqref{h3} may vanish (thus actually reaching the lower bound zero) is when
$\mathbf{p}=\mathbf{0}$ and $\mathbf{x}$ is such that
$V_j(\mathbf{x})$ is positive for $j=1,\ldots ,n$. This requires in particular that
$V_j(\mathbf{x})$ is real-valued, i.e. that
\begin{subequations}
\begin{equation}\label{bethe1}
V_j(\mathbf{x})=V_j(-\mathbf{x}), \qquad j=1,\ldots ,n,
\end{equation}
or more explicitely
\begin{gather}\label{bethe2}
\prod_{1\leq k\leq n,\, k\neq j}
\frac{\sin (x_j+x_k+ig)\sin (x_j-x_k+ig)}{\sin (x_j+x_k-ig)\sin (x_j-x_k-ig)}
= \\ \qquad \frac{\sin (x_j-ig_1)\cos (x_j-ig_2 )\sin (x_j-ig_3)\cos(x_j-ig_4)}
       {\sin (x_j+ig_1)\cos (x_j+ig_2 )\sin (x_j+ig_3)\cos(x_j+ig_4)}
       , \qquad j=1,\ldots ,n . \nonumber
\end{gather}
\end{subequations}
We will see below that the nonlinear system of algebraic equations
in Eq. \eqref{bethe2} has a unique solution $0<x_1<x_2<\cdots <x_n<\pi/2$ given by the zeros of the Askey-Wilson polynomial of degree $n$.
It is not difficult to see that for this solution in fact $V_j(\mathbf{x})>0$ for $j=1,\ldots ,n$, whence $H(\mathbf{0},\mathbf{x})=0$.
Indeed, $V_j(\mathbf{x})$ is real-valued by Eq. \eqref{bethe1}. Furthermore,
for sufficiently small
values of the coupling parameters $V_j(\mathbf{x})$ must be positive as the function in question
tends to $1$ for $g, g_r\to 0$.
This positivity remains valid for general positive parameter values $g$, $g_r$ by a continuity argument revealing that the sign cannnot flip (as none of the factors in
$V_j(\mathbf{x})$ \eqref{h2}, \eqref{h3} becomes zero or singular).

We thus arrive at the following theorem.
\begin{theorem}\label{thm1}
The trigonometric $BC$-type Ruijsenaars-Schneider Hamiltonian $H(\mathbf{p},\mathbf{x})$ \eqref{h1}-\eqref{h3}
assumes the global minimum $H=0$ only at the point in the phase space
$\Omega$ \eqref{phase-space} such that $p_1=p_2=\cdots =p_n=0$ and
$0<x_1<x_2<\cdots <x_n<\pi/2$ form a solution of the nonlinear system of algebraic equations
in Eq. \eqref{bethe2}.
\end{theorem}

\section{Zeros of the Askey-Wilson polynomials}\label{sec4}
The nonlinear system of algebraic equations in Eq. \eqref{bethe2} turns out to be
a special case
of the Bethe Ansatz equations associated to $q$-Sturm-Liouville problems
studied recently by Ismail {\it et al} \cite{ism-lin-roa:bethe}. It follows from
the machinary in {\it loc. cit.} that this algebraic system has a unique solution
given by the zeros of the Askey-Wilson polynomial of degree $n$. Below we will provide
an independent direct proof of this fact.

To this end, we first need to recall some basic properties of
the Askey-Wilson polynomials taken from Ref. \cite{ask-wil:some}.
The (monic) Askey-Wilson polynomials are trigonometric polynomials of the form
\begin{subequations}
\begin{equation}
p_n(x) = \cos(2nx)+\sum_{k=0}^{n-1} a_k \cos (2kx),\qquad n=0,1,2,\ldots ,
\end{equation}
obtained by applying Gram-Schmidt orthogonalization of the standard Fourier cosine basis
$1,\cos(2x),\cos(4x),\ldots$ on the interval $(0,\pi/2)$
with repect to the inner product
\begin{equation}
\langle f,g\rangle_\Delta =\int_0^{\pi/2}  f(x)\overline{g(x)} \Delta (x) \text{d} x ,
\end{equation}
associated to the weight function
\begin{equation}\label{weight}
\Delta (x)=\frac{1}{c(x)c(-x)}, \qquad c(x) =\frac{(ae^{2ix},b^{2ix},ce^{2ix},de^{2ix};q)_\infty}{(e^{4ix};q)_\infty}.
\end{equation}
\end{subequations}
Here it is assumed that all parameters are real-valued subject to
the constraints $0<q<1$ and $0<|a|,|b|,|c|,|d|<1$. These parameter restrictions ensure in particular that the weight function
$\Delta (x)$ \eqref{weight} is positive in the interval $(0,\pi/2)$.
The Askey-Wilson polynomials admit an explicit representation in terms of the following terminating
basic hypergeometric series
\begin{equation}\label{aw-pol}
p_n(x)=  \frac{(ab,ac,ad ;q)_n}{2\, a^n (abcdq^{n
-1};q)_n}{}_4\Phi_3
\left[ \begin{array}{c} q^{-n},abcdq^{n-1},ae^{2ix},ae^{-2ix} \\
ab,ac,ad \end{array}  \mid q;q  \right] .
\end{equation}
The polynomials under consideration are the eigenfunctions of a second-order difference operator.
Upon performing the parameter substitution
\begin{equation}\label{aw-par}
q=e^{-2g},\quad a=e^{-2g_1},\quad b=-e^{-2g_2},\quad c=e^{-2g_3},\quad d=-e^{-2g_4} ,
\end{equation}
the corresponding eigenvalue equation becomes of the form
\begin{subequations}
\begin{equation}\label{de1}
Dp_n(x)=E_n p_n(x),
\end{equation}
where $D$ denotes the difference operator
\begin{equation}\label{de2}
D= W(x)(T_{ig}-1)+W(-x)(T_{-ig}-1)\qquad ( (T_{ig}f)(x):=f(x+ig) ),
\end{equation}
with
\begin{equation}\label{de3}
W(x)= \frac{\sin (x+ig_1)\cos (x+ig_2)\sin (x+ig_3)\cos(x+ig_4)}{\sin(2x)\sin(2x+ig)} ,
\end{equation}
and the eigenvalue is given by
\begin{equation}\label{de4}
E_n=\bigl(  \cosh (\hat{g}+2ng )-\cosh ( \hat{g} ) \bigr) /2,\qquad
\hat{g}=g_1+g_2+g_3+g_4-g .
\end{equation}
\end{subequations}

After these preliminaries, we are now in the position to prove the main result.
It follows from the general fact that the polynomials form an orthogonal system
on the interval $(0,\pi/2)$ with respect to a positive weight function that the Askey-Wilson polynomial $p_n(x)$ has $n$ simple zeros inside the interval $(0,\pi/2)$. If we denote these zeros
by $x_1,\ldots ,x_n$, then it is clear that the Askey-Wilson polynomial factorizes as
\begin{equation}\label{aw-fac}
p_n(x)=2^{2n-1} \prod_{k=1}^n \sin (x_k+x)\sin (x_k-x)
\end{equation}
(since $2\sin (x_k+x)\sin (x_k-x)=\cos(2x)-\cos(2x_k)$).
After plugging the factorization of the Askey-Wilson polynomial from Eq. \eqref{aw-fac} into the difference equation in Eqs. \eqref{de1}-\eqref{de4}, and setting of $x$ equal to the $jth$ root
$x_j$, one arrives at the identity
\begin{gather}\label{vanish}
W(x_j)\prod_{k=1}^n \sin (x_k+x_j+ig)\sin (x_k-x_j-ig)\; + \\
W(-x_j)\prod_{k=1}^n \sin (x_k+x_j-ig)\sin (x_k-x_j+ig)=0,  \nonumber
\end{gather}
which amounts to Eq. \eqref{bethe2}. This shows that the roots of the Askey-Wilson polynomial
solve the nonlinear system of algebraic equations in Eq. \eqref{bethe2}.

To see that this is the only solution (up to permutation), we now assume---reversely---that the points $0<x_k<\pi/2$, $k=1,\ldots ,n$ are such that they constitute {\em any} solution to Eq. \eqref{bethe2}, and show that this implies that the corresponding factorized polynomial of the form $p_n(x)$ \eqref{aw-fac} must be equal to the Askey-Wilson polynomial. To this end it is sufficient to infer
that the factorized polynomial in question solves the eigenvalue equation in Eqs. \eqref{de1}-\eqref{de4} (since the spectrum of $D$ is nondegenerate as a continuous function of the parameters and thus determines the eigenpolynomials uniquely).
It is clear from the fact that the Askey-Wilson polynomials form the corresponding eigenbasis that acting with the operator $D$ \eqref{de2}, \eqref{de3} on a monic polynomial of the form in
Eq. \eqref{aw-fac} produces the eigenvalue $E_n$ \eqref{de4} times a certain monic polynomial $q_n(x)$ of degree $n$. Furthermore, we have that $E_nq_n(x_j)=(Dp_n)(x_j)=0$ for $j=1,\ldots ,n$, because
of Eq. \eqref{vanish} (which holds since the points $x_1,\ldots ,x_n$ solve Eq. \eqref{bethe2} by assumption).
Hence, the monic polynomial $q_n(x)$ has the same roots as $p_n(x)$ and thus coincides with it.
In other words,
the factorized polynomial $p_n(x)$ solves the Askey-Wilson difference equation, and is thus equal to
the Askey-Wilson polynomial, whence the roots $x_1,\ldots ,x_n$ correspond to the roots of the Askey-Wilson polynomial. This gives rise to the following theorem.

\begin{theorem}\label{thm2}
The unique (up to permutation) solution $0<x_k<\pi/2$, $k=1,\ldots ,n$ of the nonlinear system of algebraic
equations in Eq. \eqref{bethe2} is given by the (simple) roots of the Askey-Wilson polynomial
$p_n(x)$ \eqref{aw-pol} with parameters of the form in Eq. \eqref{aw-par}.
\end{theorem}
By combining Theorem \ref{thm1} and Theorem \ref{thm2}, we end up with the desired charaterization of
the ground-state equilibrium configuration in terms of zeros of the Askey-Wilson polynomial.
\begin{corollary}
The trigonometric $BC$-type Ruijsenaars-Schneider Hamiltonian $H(\mathbf{p},\mathbf{x})$ in Eqs. \eqref{h1}-\eqref{h3}
assumes the global minimum $H=0$ only at the point in the phase space
$\Omega$ \eqref{phase-space} such that $p_1=p_2=\cdots =p_n=0$ and
$0<x_1<x_2<\cdots <x_n<\pi/2$ are given by the (simple) roots of the Askey-Wilson polynomial
$p_n(x)$ \eqref{aw-pol} with parameters of the form in Eq. \eqref{aw-par}.
\end{corollary}

\section{Rational degeneration}\label{sec5}
By working the way down the Askey hierarchy of (basic) hypergeometric orthogonal polynomials, starting from the Askey-Wilson polynomials corresponding to the trigonometric $BC$-type Ruijsenaars-Schneider systems,
one arrives at the equilibrium configurations associated to the degenerate Ruijsenaars-Schneider systems considered by Sasaki {\it et al} \cite{rag-sas:quantum,oda-sas:equilibria} and at the equilibrium configurations associated to the Calogero-Moser systems considered by Calogero {\it et al} \cite{cal:zeros,cal:equilibrium,cal-per:properties,ols-per:classical}.
As an example, we will wrap up by detailing the important case of the rational $BC$-type Ruijsenaars-Schneider system \cite{die:integrability}. In this case the ground-state equilibrium turns out to be given by the zeros of the Wilson polynomials \cite{oda-sas:equilibria}.

The Hamiltonian $H(\mathbf{p},\mathbf{x})$ of the rational $BC$-type Ruijsenaars-Schneider system is given by Eqs. \eqref{h1}, \eqref{h2} with potentials of the form \cite{die:integrability}
\begin{equation}\label{hw}
v(x)= (x+ig)/x,\quad
w(x)= (x+ig_1)(x+ig_2)(x+ig_3)(x+ig_4)/x^2 .
\end{equation}
The phase space becomes in this situation
\begin{equation}\label{phasew}
\Omega =
\{ (\mathbf{p},\mathbf{x})\in \mathbb{R}^{2n}\mid
0< x_1 <x_2<\cdots < x_{n-1}<x_n   \} .
\end{equation}

The following theorem characterizes the ground-state equilibrium configuration of the rational $BC$-type Ruijsenaars-Schneider Hamiltonian in terms of the zeros of the Wilson polynomials
\cite{wil:some,koe-swa:askey}
\begin{subequations}
\begin{gather}\label{w-pol}
p_n(x)= \frac{(-1)^n (a+b,a+c,a+d)_n}{(n+a+b+c+d-1)_n} \\
\qquad \quad \times \; {}_4F_3
\left[ \begin{array}{c} -n,n+a+b+c+d-1,a+ix,a-ix \\
a+b,a+c,a+d \end{array}  \mid 1 \right] , \nonumber
\end{gather}
which satisfy the orthogonality relations
\begin{equation}
\int_0^\infty p_n(x)\overline{p_m(x)} \Delta (x) \text{d}x =0,\qquad n\neq m ,
\end{equation}
associated to the positive weight function
\begin{equation}
\Delta (x)=\frac{1}{c(x),c(-x)},
\quad c(x)=\frac{\Gamma (2ix)}
                {\Gamma (a+ix)\Gamma (b+ix)\Gamma (c+ix)\Gamma (d+ix)} ,
\end{equation}
\end{subequations}
where $a,b,c,d>0$ (and $\Gamma (\cdot)$ refers to the gamma function).

\begin{theorem}
The rational $BC$-type Ruijsenaar-Schneider Hamiltonian $H(\mathbf{p},\mathbf{x})$ from Eqs. \eqref{h1},\eqref{h2}, with potentials of the form in Eq. \eqref{hw},
has a unique global minimum $H=0$ in the phase space $\Omega$ \eqref{phasew} at
$p_1=p_2=\cdots =p_n=0$ and
$0<x_1<x_2<\cdots <x_n$ given by the (simple) roots of the rescaled Wilson polynomial
$p_n(x/g)$ \eqref{w-pol} with rescaled parameters $a=g_1/g$, $b=g_2/g$, $c=g_3/g$, $d=g_4/g$; these roots in turn
constitute the
unique positive solution to the nonlinear system of algebraic equations
\begin{equation*}
\prod_{1\leq k\leq n,\, k\neq j}
\frac{(x_j+x_k+ig)(x_j-x_k+ig)}{ (x_j+x_k-ig) (x_j-x_k-ig)}
= \frac{ (x_j-ig_1) (x_j-ig_2 ) (x_j-ig_3) (x_j-ig_4)}
       { (x_j+ig_1) (x_j+ig_2 ) (x_j+ig_3) (x_j+ig_4)} ,
\end{equation*}
$j=1,\ldots ,n$.
\end{theorem}
It is clear from the theorem that varying the value of the coupling parameter $g$ gives rise to a linear rescaling of the equilibrium positions.
More specifically, starting from the $g=1$ configuration corresponding to the zeros $x_1,\ldots ,x_n$
of the Wilson polynomials
$p_n(x)$ \eqref{w-pol} with $a=g_1$, $b=g_2$, $c=g_3$, $d=g_4$, one
passes to the equilibrium configuration for general positive $g$ via the rescaling
$x_j\to gx_j$, $j=1,\ldots ,n$.

The proof of the above theorem runs along the same lines of Sections \ref{sec3} and \ref{sec4},
and hinges on the second-order difference
equation for the rescaled Wilson polynomials
of the form in Eqs. \eqref{de1}, \eqref{de2} with
\begin{equation}
W(x)= \frac{(x+ig_1)(x+ig_2)(x+ig_3)(x+ig_4)}{2x(2x+ig)} ,\qquad
E_n=-ng(ng+\hat{g})
\end{equation}
(cf. e.g. \cite{koe-swa:askey}). Indeed, it is immediate that the minimization condition $V_j(\mathbf{x})=V_j(-\mathbf{x})$ now gives rise to the algebraic equations for the equilibrium points stated in the theorem; furthermore,
the solution of this system readily follows upon substitution of the factorization
$p_n(x)=\prod_{k=1}^n (x+x_k)(x-x_k)$ into the difference equation for the rescaled
Wilson polynomials.

\section*{Acknowledgements}
This work was supported in part by the Fondo
Nacional de Desarrollo Cient\'{\i}fico y Tecnol\'ogico (FONDECYT)
Grant No. \# 1010217.

\label{lastpage}

\end{document}